\documentclass[conference]{IEEEtran}
\IEEEoverridecommandlockouts
\usepackage[utf8]{inputenc}
\usepackage[T1]{fontenc}
\usepackage{cite}
\usepackage{amsmath,amssymb,amsfonts}
\usepackage{graphicx}
\usepackage{textcomp}
\usepackage{xcolor}
\usepackage{comment}
\usepackage[hidelinks]{hyperref}
\def\BibTeX{{\rm B\kern-.05em{\sc i\kern-.025em b}\kern-.08em
    T\kern-.1667em\lower.7ex\hbox{E}\kern-.125emX}}

\newtheorem{theorem}{Theorem}

\begin{document}

\title{Harmoniq: Efficient Data Augmentation on a Quantum Computer Inspired by Harmonic Analysis
}

\author{\IEEEauthorblockN{Kristina Kirova}
\IEEEauthorblockA{\textit{Dept.\ for Quantum Information} \\
\textit{and Computation (QUICK)} \\
\textit{Johannes Kepler University}\\
Linz, Austria \\
kristina.kirova@jku.at}
\and
\IEEEauthorblockN{Monika D{\"o}rfler}
\IEEEauthorblockA{\textit{Department of Mathematics} \\
\textit{University of Vienna}\\
Vienna, Austria \\
monika.doerfler@univie.ac.at}
\and
\IEEEauthorblockN{Franz Luef}
\IEEEauthorblockA{\textit{Department of} \\
\textit{Mathematical Sciences} \\
\textit{Norwegian University of} \\
\textit{Science and Technology}\\
Trondheim, Norway \\
franz.luef@ntnu.no}
\and
\IEEEauthorblockN{Richard Kueng}
\IEEEauthorblockA{\textit{Dept.\ for Quantum Information} \\
\textit{and Computation (QUICK)} \\
\textit{Johannes Kepler University}\\
Linz, Austria \\
richard.kueng@jku.at}
}

\maketitle
\begin{abstract}
Quantum machine learning has attracted significant interest in recent years. Most existing approaches, however, are variational in nature and require extensive parameter optimization subroutines.
Here, we propose a conceptually distinct quantum machine learning approach that goes beyond the variational paradigm.
Harmoniq takes a recently developed data augmentation technique from quantum harmonic analysis and implements it as a stochastic mixture of n-qubit circuits with at most quadratic depth each. 
A key strength of Harmoniq is its modularity: viewed as a quantum process acting on density matrices, it can readily be combined with other quantum data processing and learning subroutines. A subsequent case study demonstrates this modularity by combining Harmoniq with stochastic amplitude encoding for the input density matrix and quantum PCA on the output density matrix. This results in a promising signal denoising pipeline that works particularly well in the small sample size regime.
\end{abstract}

\begin{figure*}
  \begin{centering}
  \includegraphics[width=1.0\linewidth]{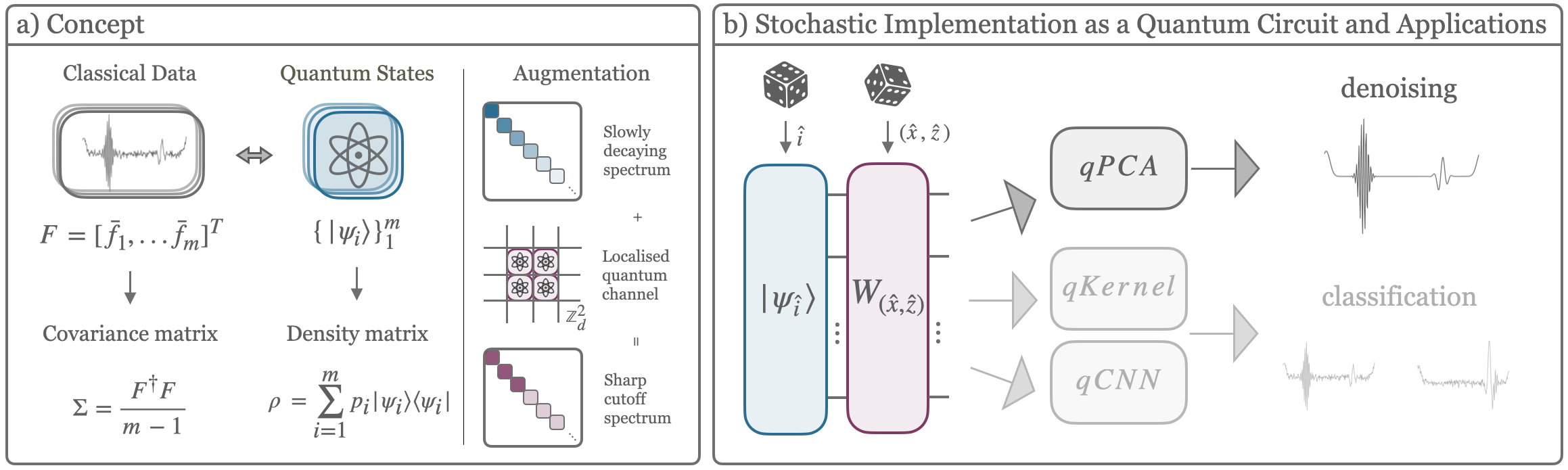}
  \caption{\emph{Concept and Implementation of \emph{Harmoniq}.}
  (a) We take a set of classical data signals and normalize it to a set of pure quantum states. With this, the covariance matrix of the data can be interpreted as a mixed quantum state (also known as a density matrix). The augmentation via the localised Weyl-Heisenberg quantum channel results in an eigenvalue spectrum with a sharper cutoff.
  (b) The quantum channel can be implemented stochastically with an efficient quantum circuit by drawing pure quantum states and Weyl-Heisenberg matrices from their underlying distributions. Here, we focus on one specific application (denoising) but many others are possible.}
  \label{fig:main_fig}
\end{centering}
\end{figure*}

\section{Introduction}
Recent years have seen growing interest in the use of quantum computers for machine learning tasks. The central motivation is that quantum systems naturally process information in high-dimensional Hilbert spaces and can implement transformations that may be difficult to simulate classically. This has led to the emergence of quantum machine learning (QML)~\cite{schuld2015introduction, 
schuld2018supervised, 
richard_experiments_2022}, a field that explores how quantum resources can be used for data analysis and pattern recognition. Several distinct paradigms for quantum machine learning have emerged, including quantum kernel methods~\cite{schuld2019feature, havlicek2019supervised} and variational quantum neural networks~\cite{farhi2018classification, cerezo2021variational}.

Many proposed QML approaches draw inspiration from successful classical machine learning architectures and translate them to the quantum realm. For example, quantum support vector machines (qSVMs)~\cite{qSVM}, quantum convolutional neural networks (qCNNs)~\cite{qCNN} and quantum generative adversarial networks (qGANs)~\cite{qGAN} arose in this fashion. While this strategy has produced a variety of hybrid quantum–classical models, several works have highlighted fundamental challenges in their trainability and scalability, including barren plateau phenomena and limits on achievable quantum advantage~\cite{mcclean2018barren, cerezo2021variational}. Moreover, most research in QML has focused on the design of learning models and circuit architectures, while comparatively little attention has been devoted to data-centric techniques. In classical machine learning, however, methods such as data augmentation play a central role in improving generalization by incorporating prior knowledge about the structure of the data~\cite{shorten2019survey, cubuk2019autoaugment}. Systematic strategies for augmenting datasets via quantum circuits remain largely unexplored. These observations motivate the search for mathematically grounded approaches to quantum data processing.

Quantum harmonic analysis (QHA)~\cite{Werner84} provides one such framework. Building on ideas from classical harmonic analysis, QHA systematically extends tools such as (standard and short-time) Fourier transforms, time-frequency analysis~\cite{book_TF} and related spectral techniques to the non-commutative setting of quantum mechanics~\cite{dorfler_local_2024, doerfler2025}. Rather than heuristically adapting classical algorithms, this approach offers mathematically well-founded procedures for processing quantum data and defining transformations directly at the level of quantum states, evolutions and measurements.

In this work, we introduce \emph{Harmoniq}: a quantum data augmentation protocol derived from quantum harmonic analysis. \emph{Harmoniq} is fully analytical, contains no trainable parameters, and does not rely on optimization or kernel methods. Instead of learning from the data, the algorithm applies well-defined transformations to enhance latent structure already present within the data. Because the channel is defined independently of the downstream task, it serves as a modular building block: it can precede quantum PCA, quantum kernel methods, or qCNNs, operating throughout on data already encoded in the register.

We demonstrate the framework in the context of signal denoising, where we combine it with quantum principal component analysis, and show that it is particularly effective in the small-sample regime. For this specific pipeline we do not claim an end-to-end speedup as amplitude encoding and the copy requirements of qPCA dominate the overall cost. The augmentation layer itself, however, is exponentially cheaper in memory and depth than applying the same convolution to an explicit $2^n \times 2^n$ covariance matrix.

\section{Background}
This section gives a brief overview of the most important technical concepts which this work relies on, namely quantum harmonic analysis, Weyl-Heisenberg matrices and dimensionality reduction (PCA). It also establishes the formal link between signal processing in the time-frequency domain and the algebraic structure of finite-dimensional quantum systems.

\subsection{Quantum Harmonic Analysis (QHA)}
In classical harmonic analysis, the study of signals typically takes place in the continuous Hilbert space $L^2(\mathbb{R})$ 
of square-integrable functions in one real-valued variable, where the fundamental building blocks are the time-shift $T_x$ and frequency-shift $M_z$ operators. For a function $g \in L^2(\mathbb{R})$, translation by $x$ acts as $T_x g(t) = g(t-x)$, while a frequency shift by $z$ operates as $M_z g(t) = \mathrm{e}^{2\pi \mathrm{i} z t} g(t)$. The composition of both operators yields a \emph{time-frequency (TF) shift} operator labeled by $(x,z) \in \mathbb{R}^2$:
\begin{equation}\label{eq:TF-QHA}
W(x,z) g(t) = M_z T_x g(t) = \mathrm{e}^{2\pi \mathrm{i} z t} g(t-x) \end{equation}
for all $g \in L^2 (\mathbb{R})$.

While classical analysis focuses on functions, QHA, as pioneered by Werner~\cite{Werner84} in 1984, lifts these concepts to the level of operators.  Central notions of classical harmonic analysis, in particular translations and convolutions, are introduced on an operator level in order to transfer their analytic properties to the operator setting. Much like unitary evolutions affect density matrices by conjugation, operator-valued time-frequency shifts map a trace-class operator $O$ to $\alpha_{(x,z)}(O) = W(x,z) O W(x,z)^\dagger$. Based on this action, convolution between  a trace-class operator $O$ and an integrable function  $\lambda$ on $\mathbb{R}^2$ is defined as the operator
\begin{align}\label{eq:def_conv}
\lambda\star O
=&\int_{\mathbb{R}^2} \lambda_{(x,z)} \alpha_{(x,z)}(O) \mathrm{d}x \mathrm{d}z \nonumber \\
=& \int_{\mathbb{R}^2} \lambda_{(x,z)} W_{(x,z)} O W_{(x,z)}^\dagger \mathrm{d}x \mathrm{d}z
\end{align}

For our purposes, the key property of this convolution is that it acts as a smoothing operation on the operator $O$. In harmonic analysis terms, $\lambda \star O$ is obtained by integrating over translated copies of $O$ across different time-frequency shifts, with weights given by $\lambda_{(x,z)}$. With this, fine-scale variations in $O$ and unstructured noise are suppressed, yielding a more regular operator.

This behavior has a direct and powerful application in data science. When $O$ is interpreted as the covariance matrix of a dataset, the convolution $\lambda \star O$ represents the covariance of an augmented version of that data. As shown in Ref.~\cite{doerfler2025}, this increased regularity translates to a faster decay in the operator's eigenvalues. In the context of Principal Component Analysis (cf.~Section~\ref{sec:PCA}), this means that the augmentation effectively yields a sharper spectral cutoff, as illustrated in Fig.~\ref{fig:main_fig}~(a). 

While the theoretical foundations of QHA are typically framed in the continuous setting, all constructions carry over exactly to the finite-dimensional Hilbert space used in quantum computing, as we elaborate in the following subsection.

\subsection{Weyl-Heisenberg Matrices}
In the context of quantum computing, one typically focuses on a $d$-dimensional Hilbert space $\mathbb{C}^d$, where $d=2^n$ and $n$ denotes the number of qubits. Over the finite phase space $\mathbb{Z}_d \times \mathbb{Z}_d$, the time-frequency shift operators are the Weyl–Heisenberg matrices (also called generalized Pauli operators).

Every Weyl-Heisenberg matrix is specified by a discrete parameter pair $(x,z) \in \left\{0,\ldots,d-1\right\}^{\times 2}$ and assumes the following form:
\begin{equation} \label{eq:WH}
W(x,z) = \mathrm{e}^{- \mathrm{i}\pi xz/d} Z_d^z X_d^x,
\end{equation}
where
\begin{align*}
    Z_d =& \sum_{k=0}^{d-1} \omega_d^k |k \rangle \! \langle k| & \text{(clock, generalized $Z$)}, \\
    X_d =& \sum_{k=0}^{d-1} |k +1  \bmod d\rangle \! \langle k| & \text{(shift, generalized $X$)},
\end{align*}
and $\omega_d = \exp \left( 2 \pi \mathrm{i}/d \right)$ is a $d$-th root of unity.

These matrices feature prominently in many aspects of quantum information science, such as qudit-based quantum information~\cite{Gibbons2004_phase_space, Ringbauer_2022_qudits, brandl2026quickquditsframeworkefficientsimulation}, stabilizer theory~\cite{Gottesman1997_codes, Gross2006, kueng2015qubitstabilizerstatescomplex} and optimal measurements~\cite{Lawrence2002_MUBs}.
Here, we identify them as the discrete version of the time-frequency shift operator from \eqref{eq:TF-QHA}. This allows us to draw a connection to QHA.
In particular, the convolution from \eqref{eq:def_conv} can be interpreted as a quantum channel acting on a particular trace-class operator, namely the density matrix describing a quantum state, $\rho$, as
\begin{equation}\label{eq:quantum-channel}
\mathcal{P}_\Lambda(\rho)=\sum_{{(x,z)}}\lambda_{(x,z)} W_{(x,z)}\rho W_{(x,z)}^\dagger,
\end{equation}
where $\Lambda = \{\lambda_{(x,z)}\}$ is the collection of
weights, with $\lambda_{(x,z)} \ge 0$ and $\sum_{(x,z)}\lambda_{(x,z)} = 1$. Operationally, this channel corresponds to a probabilistic mixture of $d^2$ unitary evolutions, where each unitary is a Weyl-Heisenberg matrix $W_{(x,z)}$ and the associated probability weight is $\lambda_{(x,z)}$.
This is the foundation of our work, which brings fundamental concepts from harmonic analysis to the finite-dimensional setting of quantum computing, to propose an efficient algorithm for data augmentation.

\subsection{Principal Component Analysis} \label{sec:PCA}
Principal Component Analysis (PCA) is a dimensionality-reduction technique that identifies the directions (principal components) along which a given dataset varies the most~\cite{jolliffe2002principal}. 
Suppose that we have $m$ data points $f_i \in \mathbb{C}^d$ in $d$ dimensions. We first center the data by subtracting the empirical mean, $f_i \mapsto \bar{f}_i = f_i - \frac{1}{m} \sum_{j=1}^m f_j$, and collect the centered data vectors as (conjugated) rows in the data matrix $F = \left(\bar{f}_1 \cdots \bar{f}_m\right)^\dagger\in \mathbb{C}^{m \times d}$. PCA then diagonalizes the associated covariance matrix $\Sigma = \frac{1}{m-1} F^\dagger F$ and finds its eigenvectors and eigenvalues. The eigenvectors define orthogonal directions in feature space, and the eigenvalues quantify the variance captured along each direction. By projecting the data onto the first $K$ eigenvectors, one obtains a lower-dimensional representation that preserves as much variance as possible~\cite{shlens2014tutorial}. PCA is widely used for data compression, noise reduction, visualization, and as a preprocessing step for machine learning.
\subsection{Quantum Principal Component Analysis}
Quantum Principal Component Analysis (qPCA) is a quantum algorithmic framework designed to extract the principal components of a density matrix or covariance operator using quantum resources. In a seminal work by Lloyd, Mohseni, and Rebentrost \cite{qPCA}, qPCA is formulated in terms of density matrix exponentiation: given access to multiple copies of a quantum state $\rho$, one can efficiently simulate the unitary evolution $\mathrm{e}^{-\mathrm{i} \rho t}$ and apply quantum phase estimation to obtain its eigenvalues and eigenvectors. The algorithm requires order $\mathcal{O}(1/\epsilon^3)$ copies of $\rho$ to estimate eigenvalues to precision $\epsilon$, and the gate complexity scales linearly with the number of qubits, $n$. Since $n$  qubits encode a $d=2^n$ dimensional vector, qPCA yields an exponential speed-up over classical diagonalization techniques. Variational approaches to qPCA have also been proposed~\cite{VarPCA}, where a parametrized quantum circuit is trained to approximate the leading eigenspace of $\rho$, reducing resource requirements on near-term devices.
\section{Framework}\label{framework}
In this section, we present the core of \emph{Harmoniq}: an informed data augmentation procedure which can be efficiently executed on a quantum computer. Using the tools of quantum harmonic analysis, we interpret the augmentation of a dataset as a localized quantum channel acting on a density matrix that encodes the data's statistical structure.

Let $F \in \mathbb{C}^{m \times d}$ denote the data matrix whose (conjugated) rows are the centered data vectors $\bar{f}_i \in \mathbb{C}^d$. Studying the eigenspectrum of the data covariance matrix $\Sigma=\frac{1}{m-1} F^\dagger F$ reveals the underlying structure in the data. This is crucial for many data science applications, most notably PCA, as outlined in Sec.~\ref{sec:PCA}. The normalized data covariance matrix $\Sigma/\mathrm{tr}(\Sigma)$ can be encoded as a mixed $d$-dimensional quantum state 
\begin{equation} \label{eq:density-matrix}
    \rho=\sum_{i=1}^m p_i |\psi_i\rangle\langle \psi_i|,
\end{equation} 
where $|\psi_i \rangle$ is the quantum state corresponding to an amplitude encoding of the normalized $\bar{f}_i$, $|\psi_i \rangle = \bar{f}_i /||\bar{f}_i||_2$ and every vector enters with a probability proportional to its squared norm, $p_i = \| \bar{f}_i\|_2^2 / \sum_{j=1}^{m} \|\bar{f}_j\|_2^2$. This correspondence is not new and has, for instance, been used in Ref.~\cite{CovMatrix_quantum_LosAlamos}. Operationally, the density matrix in \eqref{eq:density-matrix} admits a stochastic realization: it can be viewed as the ensemble average over pure states obtained by randomly sampling data points according to their energy (squared norm normalized) and encoding them as quantum states,
\begin{align}
|\hat{\psi} \rangle =  \bar f_{\hat{i}}/\|\bar f_{\hat{i}}\|_2 \quad \text{with} \quad \mathrm{Pr} \left[\hat{i}=i \right] = p_i.
\label{eq:state-realization}
\end{align}

We now transform this mixed quantum state using the Weyl-Heisenberg (generalized Pauli) channel defined in \eqref{eq:quantum-channel}. This channel randomly applies phase-space displacements to the quantum state, which are sampled from a specific distribution. This resembles an operator-function convolution in the language of quantum harmonic analysis \eqref{eq:def_conv} and a covariance-level data augmentation in the language of data science. Crucially, we do not implement the channel for the uniform distribution over all possible Weyl-Heisenberg unitaries ($\lambda_{(x,z)}=1/d^2$), as this would twirl any $\rho$ into a maximally mixed state $\mathbb{I}/d$. Instead, we define an \emph{augmentation window} $\Omega \subseteq \mathbb{Z}_{d}^2$, $|\Omega| \ll d^{2}$ specifying the (very small) subset of Weyl-Heisenberg displacements used in the protocol. This is motivated by the principle of \textit{localized augmentation} \cite{dorfler_local_2024, doerfler2025}. We elaborate further on the exact choice of augmentation in Sec.~\ref{sec:application}. The entire framework is illustrated conceptually in Fig.~\ref{fig:main_fig}~(a).

\section{Implementation on a Quantum Computer}
From now on, we set $d=2^n$ to ensure that we can implement everything on a $n$-qubit quantum computer. 
The full protocol requires data embedding, an implementation of the augmentation, and an application layer. The following subsections elaborate on one step each. A particular emphasis is put on the efficient application of the Weyl-Heisenberg channel which executes the data augmentation step. 

\subsection{Data embedding}
Data embedding is a fundamental component of quantum machine learning, where classical data is encoded into quantum states. Here, we employ amplitude encoding as it naturally maps data vectors to pure quantum states and data sets to mixed states, see \eqref{eq:density-matrix} and also Ref.~\cite{CovMatrix_quantum_LosAlamos}. While preparing arbitrary amplitude-encoded states is generally exponentially costly \cite{aaronson2015read, Plesch_2011}, this overhead can be bypassed for quantum data or mitigated via QRAM \cite{Giovannetti_2008_QRAM} and structured data loading \cite{schuld2018supervised}. Consequently, state preparation is typically treated as a black box in the algorithmic analysis, and its resource cost is discussed separately.

\begin{figure}[!t]
    \centering
    \includegraphics[width=0.9\columnwidth]{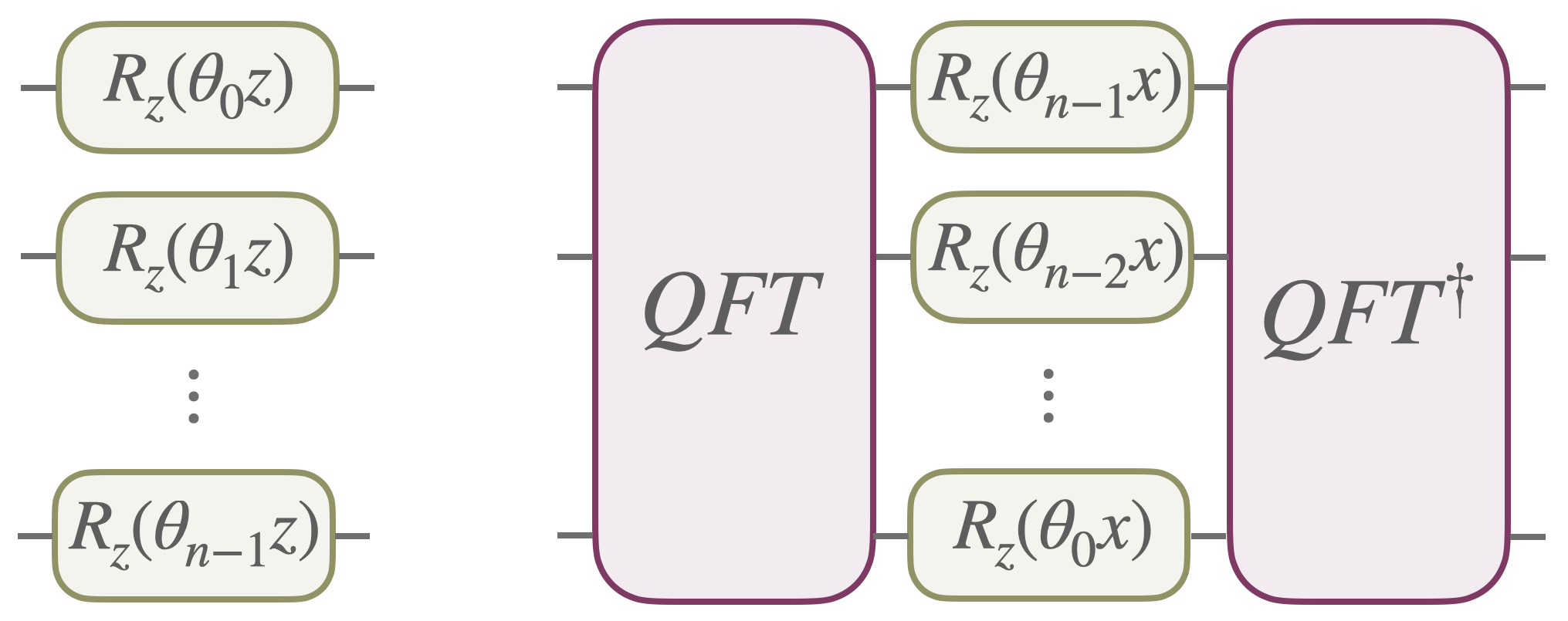} 
    \caption{Quantum circuit realizations of the generalized $Z_{d=2^n}^z$ gate (left) and $X_{d=2^n}^x$ gate (right). An arbitrary Weyl-Heisenberg matrix (for $d=2^n$) is implemented as a combination of both, according to \eqref{eq:WH} (ignoring phases).}
    \label{fig:zx_circuit}
\end{figure}

\subsection{Efficient stochastic implementation of the data augmentation channel}

Here, we show how to efficiently implement the data augmentation channel presented in \eqref{eq:quantum-channel} on a quantum computer.
This is also the core of the proposed protocol, \emph{Harmoniq}, and we refer to Fig.~\ref {fig:main_fig}(b) for a visualization of this implementation. Rather than realizing the full quantum channel deterministically, we exploit the concept of importance sampling and realize a linear combination of Weyl-Heisenberg evolutions stochastically by sampling from the underlying distribution. Namely, $\mathcal{P}_\Lambda(\rho)$ from \eqref{eq:quantum-channel} is implemented as 
\begin{equation}
    |\hat{\psi}\rangle \rightarrow W_{(\hat x, \hat z)}|\hat{\psi}\rangle \quad \text{with} \quad \text{Pr}[(\hat x, \hat z) = (x, z)] = \lambda_{(x, z)},
\end{equation}
where $|\hat{\psi} \rangle$ is the pure state sampled according to \eqref{eq:state-realization} in order to stochastically prepare $\rho$. On the level of density matrices, the unitaries act as $\rho \mapsto W_{(\hat{x},\hat{z})}\rho W_{(\hat{x},\hat{z})}^\dagger$ and taking the expectation over all possible unitaries then produces $\mathcal{P}_\Lambda (\rho)$.

The rest of this section is devoted to the realization of an arbitrary Weyl-Heisenberg matrix $W_{(x, z)}$ as an efficient quantum circuit. To the best of our knowledge, this has not been shown explicitly in a data augmentation context before.

Let us first investigate the action of the generalized $Z_d$ gate on the $|k \rangle$ state for $d=2^n$:
\begin{equation}
    Z_{2^n}|k\rangle = \mathrm{e}^{2\pi \mathrm i k/d}|k\rangle.
\end{equation}
We recognize that $k$ is an integer labeling the basis state, and we can rewrite it in a binary (qubit) basis as $k = \sum_{j=0}^{n-1} b_j 2^j$ where $j$ labels the qubit and $b_j$ is a binary number, $b_j \in \{0, 1\}$. Thus, we have:
\begin{align*}
    Z_{2^n}|b_{n-1}...b_0\rangle &= \mathrm{e}^{2\pi \mathrm i \sum_{j=0}^{n-1}b_j 2^j/d}|b_{n-1}...b_0\rangle \\
    &= \prod_{j=0}^{n-1} \mathrm{e}^{2\pi \mathrm i b_j 2^j/d}|b_{n-1}...b_0\rangle \\
    &= \bigotimes_{j=0}^{n-1} \mathrm{e}^{2 \pi \mathrm i b_j 2^j/d } |b_j \rangle \\
    &= \prod_{j=0}^{n-1} \mathrm{e}^{\mathrm i \theta_j/2} \bigotimes_{j=0}^{n-1} R_z(\theta_j)|b_j \rangle,
\end{align*}
where $\theta_j = 2\pi/2^{n-j}$ for $j \in [0, ..., n-1]$ and $R_z(\theta) = \mathrm{e}^ {- \mathrm i (\theta/2) Z} = \text{diag}(\mathrm{e}^ {- \mathrm i \theta/2}, \mathrm{e}^ {\mathrm i \theta/2})$ is the standard $R_z(\theta)$ gate. Note that this derivation adopts a computer science notation (the least significant digit is last) while in quantum computing circuits, the qubits are typically labeled in reverse order. For the latter, the angle is given by $\theta_j = 2\pi/2^{j+1}$ for $j \in [0, ..., n-1]$. 

Raising the $Z$ operator to an arbitrary power $z$  is now straightforward: just multiply the angle of all $R_z$-gates by $z$. The global phase would be multiplied accordingly:
\begin{equation}
    Z_{2^n}^z = \prod_{j=0}^{n-1} \mathrm{e}^{\mathrm i \theta_jz/2} \bigotimes_{j=0}^{n-1} R_z(\theta_jz).
\end{equation}
With this, we have shown that every Z-type Weyl-Heisenberg matrix $Z_{2^n}^z$ can be implemented in unit depth for any $d=2^n$, see Fig.~\ref{fig:zx_circuit}~(left) for an illustration. 

Now, let us turn our attention to the shift operator. Observe that $X_{2^n}^x$ implements addition mod $d$. There exists a straightforward implementation based on the fact that the quantum Fourier transform diagonalizes shift operators. By moving to the Fourier basis, the shift operator becomes a diagonal phase (clock) operator, which can be implemented efficiently, before transforming back to the computational basis via the inverse QFT, as illustrated in Fig.~\ref{fig:zx_circuit}~(right). In formulas, 
\begin{equation}
    X_{2^n} = \mathrm{QFT}^\dagger Z \mathrm{QFT}
\end{equation}
Observe that the qubit order reversal due to the QFT can be accounted for by applying the rotation angles in reversed order, and the inverse QFT restores the original ordering; see Fig.~\ref{fig:zx_circuit}~(right).

Raising the $X$ operator to a power $x$ is equivalent to raising the $Z$ operator to a power $x$, which we have already shown to be efficient. In formulas:
\begin{equation}
    X_{2^n}^x = \mathrm{QFT}^\dagger Z^x \mathrm{QFT}.
\end{equation}
Altogether, we have shown that for $d=2^n$, the implementation of an arbitrary Weyl-Heisenberg matrix can be done efficiently on a quantum computer (up to a global phase). The exact gate count is $2n$ Hadamard gates, $2n$ $R_z$ rotations, and $n(n-1)$ controlled-phase gates. The circuit depth is $\mathcal O(n^2)$, which is mainly due to the QFT blocks. We can now combine this with importance sampling to obtain an efficient stochastic implementation of the channel $\mathcal{P}_\Lambda (\rho)$ from \eqref{eq:quantum-channel}.

\begin{theorem}[Efficient implementation of the augmentation protocol]\label{theorem}
For $d=2^n$, every data augmentation channel $\mathcal{P}_{\Lambda}$ from \eqref{eq:quantum-channel} can be realized on an $n$-qubit quantum computer with a stochastic combination of unitaries of depth $\mathcal{O}(n^2)$.
\end{theorem}
To our knowledge, this is the first efficient quantum realization of the data operator convolution put forward in Ref.~\cite{dorfler_local_2024} and Ref.~\cite{doerfler2025} and displayed in \eqref{eq:def_conv}.

\subsection{Application-dependent subroutine}
Once the density matrix encoding the covariance matrix of the data has been augmented -- i.e. we mapped $\rho$ from \eqref{eq:density-matrix} to $\mathcal{P}_\Lambda (\rho)$ -- any 
downstream quantum subroutine can follow depending on the task at hand. Here, we utilize quantum PCA~\cite{qPCA}, which provides exponential speedup over classical diagonalization but requires access to $\mathcal{O}(1/\epsilon^3)$ many copies. Beyond this, we envision that \emph{Harmoniq} can enhance a broader range of learning tasks, such as classification via quantum kernel methods~\cite{schuld2019feature} or qCNNs~\cite{qCNN}.

\section{Application}\label{sec:application}
We now illustrate one application of the proposed augmentation framework, namely the denoising of data samples using (quantum) principal component analysis. 

\subsection{Dataset generation}
To evaluate the performance of \emph{Harmoniq} in a controlled setting, we construct a synthetic dataset of complex-valued signals. These signals have structured and localized features, capturing characteristics commonly observed in real-world data such as sensor measurements or spectral signals.

We begin with a discretized normalized Gaussian window $g \in \mathbb{R}^d$ of length $d = 2^n$. From it, we generate a collection of 12 ``atoms'' $\{g_k\}_{k=1}^{12}$ by applying specific time-frequency (TF) shifts:
\begin{equation}
    g_{k} = W(x_k, z_k) g.
\end{equation}
The coordinates $(x_k, z_k) \in \mathbb{Z}_d^2$ are arranged into three well-separated clusters. Each individual data vector $f_i$ is constructed as a unique realization of this underlying structure. For each signal $i = 1, \dots, m$, we sample a coefficient vector $c_i \in \mathbb{R}^{12}$ from a standard normal distribution, yielding the signal:
\begin{equation}
    f_i = \sum_{k=1}^{12} c_{k,i}\, g_k.
\end{equation}
To simulate realistic conditions, we introduce additive Gaussian noise $\epsilon \sim \mathcal{N}(0, \sigma^2)$ to both the real and the imaginary part of every entry. The addition of noise in our data set partially masks the underlying structural features of the clusters. As the noise level increases, noise masking becomes more and more pronounced which allows us to test the limits of our protocol across various noise regimes.

Following Sec.~\ref{framework}, every data vector is normalized to a quantum state and enters the probabilistic mixture proportional to its squared norm.
\begin{figure}[!t]
    \centering
    \includegraphics[width=\columnwidth]{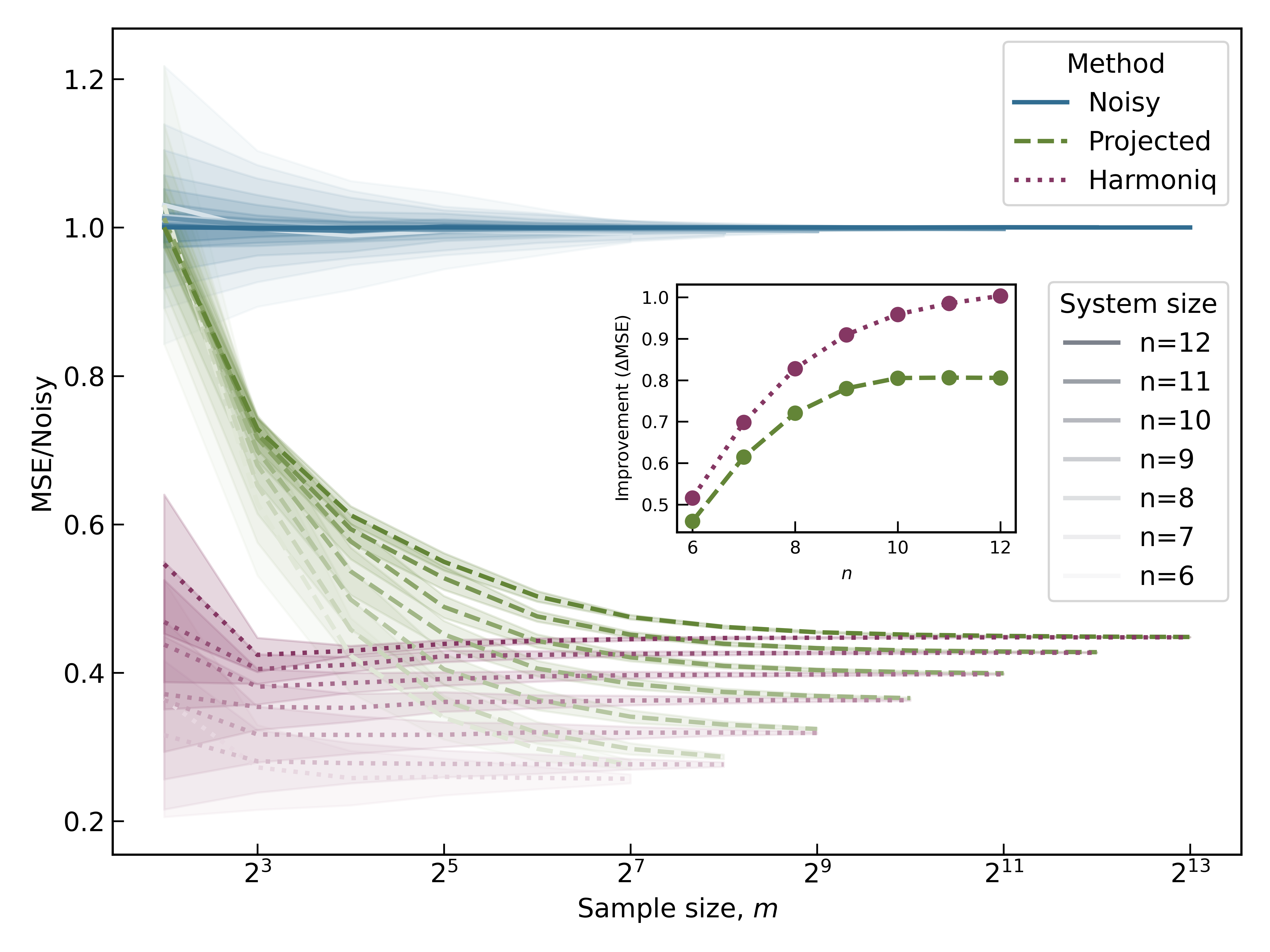} 
    \caption{\emph{Scalability of the protocol at different sample sizes.} 
    The (normalized) Mean Squared Error is significantly lower when augmenting with \emph{Harmoniq} (purple, dotted), especially when only very few samples are available. Data is averaged over $100$ instances for each system size (color gradient). Inset:
    The absolute improvement in MSE achieved achieved by projection alone and by \emph{Harmoniq}, relative to the noisy baseline increases with system size, $n$.
    }
    \label{fig:data-scarcity-with-n}
\end{figure}

\subsection{Augmentation}
The augmentation channel does not employ all Weyl-Heisenberg
matrices but restricts to an \emph{augmentation window}
$\Omega \subset \mathbb{Z}_d^2$ of phase-space displacements localized around the origin. Concretely, $\Omega$ is a centered square grid of side length $s$, with $s = n$ for odd $n$ and $s = n-1$ for even $n$, so that the size of the window grows with the number of qubits while remaining a vanishing fraction of all possible displacements.

To avoid a sharp cutoff at the window boundary, we implement a Gaussian-shaped weighting,
\begin{equation}
\lambda_{(x,z)} \propto \exp\!\left(-\frac{x^2 + z^2}{2v^2}\right),
\qquad (x,z) \in \Omega,
\end{equation}
where $v = \lfloor s/2 \rfloor /2$ and the weights are normalized to a probability distribution. Neither the choice of a window size $s$ nor the choice of $\lambda_{(x,z)}$ was tuned to the data. A systematic sweep over window parameters is deferred to future work.

\emph{Remark.} In this application, Harmoniq departs from classical
augmentation: rather than enlarging the sample
set, it acts on the second-order statistics directly. Applying the
channel to the covariance matrix is equivalent to simultaneously averaging over TF-shifted copies of every sample.

\subsection{Metric}
To quantify the benefit of data augmentation, we perform PCA on the noisy dataset with and without the augmentation. Projecting onto the largest $K$ components gives
\begin{equation}
    |\psi_i\rangle_K = \sum_{j=1}^{K} \langle v_j | \psi_i \rangle |v_j\rangle .
\end{equation}

In the section below, we refer to the projection of the non-augmented dataset simply as ``Projected'', while the projection of the augmented dataset is labeled \emph{Harmoniq}.

We measure the difference between the projected (denoised) vector and the original (clean) one using the squared $\ell_2$ norm. Averaging over all samples yields the Mean Squared Error (MSE):
\begin{equation}
    \text{MSE} = \frac{1}{m} \sum_{i=1}^{m} \| |\psi_i\rangle_K - |\psi_i\rangle_{\text{clean}} \|_2^2 .
\end{equation}
\begin{figure}[!t]
    \centering
    \includegraphics[width=\columnwidth]{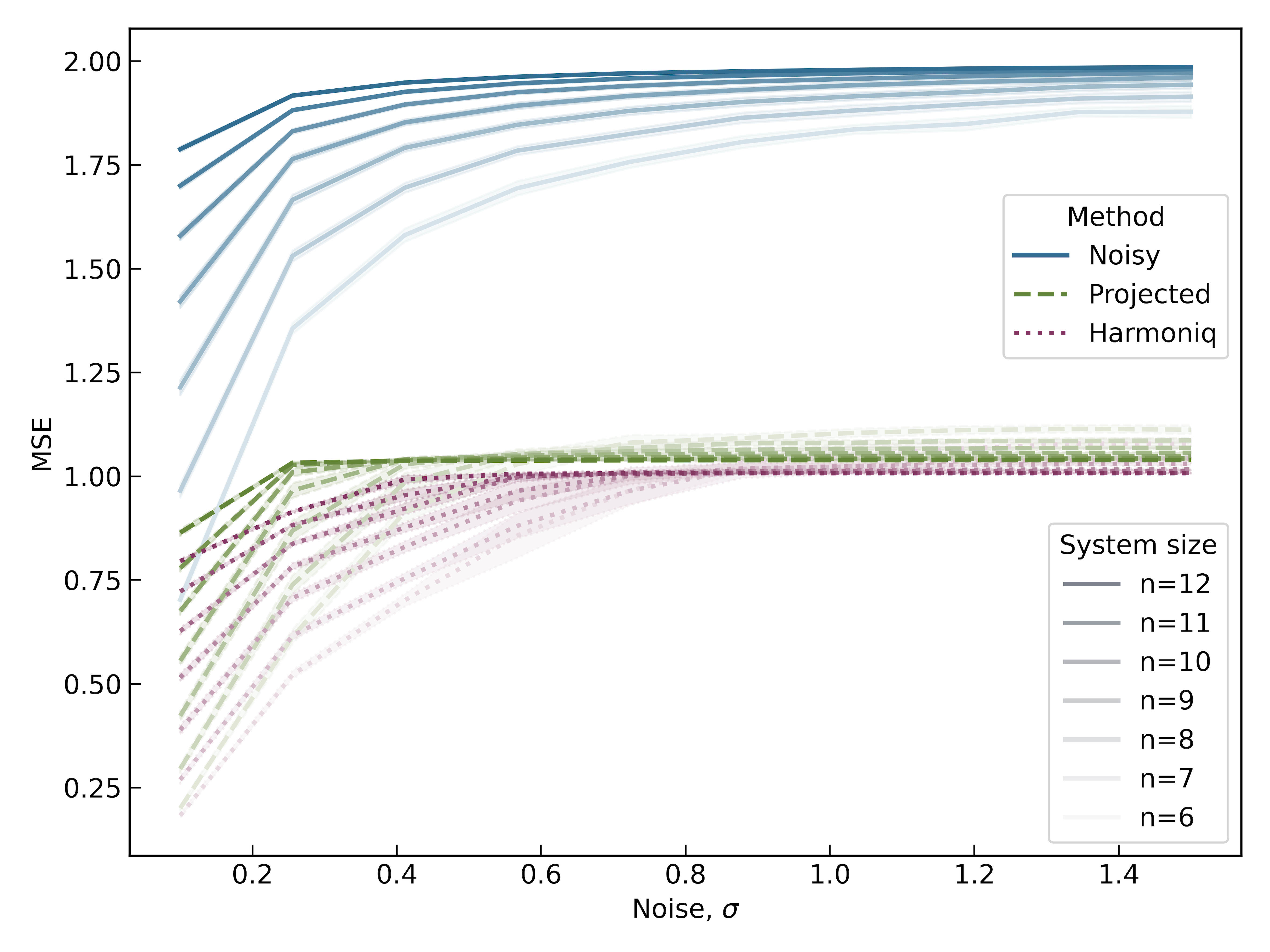} 
    \caption{\emph{Scalability of the protocol at different noise levels.} The Mean Squared Error for \emph{Harmoniq} (purple, dotted) is lower than the Projected for any noise regime. As the noise increases, performance eventually saturates near $2$.
    Results show $100$ runs per system size (color gradient) at a fixed sample size, $m=100$.}
    \label{fig:noise-levels-with-n}
\end{figure}

\subsection{Results}
We conducted two sets of numerical experiments to evaluate the impact of the augmentation for different settings. First, we changed the sample size, $m$, at a fixed noise level $\sigma = 0.1$, see Fig.~\ref{fig:data-scarcity-with-n}. Second, we varied the noise level, $\sigma$ at a fixed sample size $m=100$, see Fig.~\ref{fig:noise-levels-with-n}. In both cases, we scaled the system size from $n=6$ to $n=12$ qubits to observe how the protocol scales with the dimension. Also, we averaged over $100$ instances ($10$ noise realizations $\times$ $10$ data batches).

In the first study, we observed that performing data augmentation with \emph{Harmoniq} before projecting on the first $K=3$ eigenvectors results in vectors which are much closer to the original, clean data. This advantage is most pronounced in the low-sample regime, where classical covariance estimates are typically least reliable, and is consistent across all system sizes. This suggests that the protocol successfully captures the underlying signal structure even as the available data becomes increasingly sparse relative to the total Hilbert space volume. 

To ease visualization, we have divided every line in Fig.~\ref{fig:data-scarcity-with-n} by the mean of the Noisy MSE baseline. Under this normalization, the relative advantage gained with \emph{Harmoniq} appears to be slightly decreasing with system size, $n$. However, the absolute difference is actually increasing, as shown in the inset of Fig.~\ref{fig:data-scarcity-with-n}. We consider this absolute gap more meaningful as it represents the total magnitude of noise suppressed.

In our second experiment, we evaluated the protocol at different noise levels, $\sigma \in [0.1, 1.5]$, while fixing the sample size at $m=100$. This is motivated by a common challenge of analyzing data when available observations are strictly limited. As expected, at extreme noise levels, the signal is entirely obscured and the MSE for the noisy dataset saturates near $2$. However, in the moderate noise regime ($\sigma \in [0.1, 0.5]$), \emph{Harmoniq} provides a distinct performance gain. This is the most practically relevant window for the algorithm, as it addresses cases where the data is imperfect but the underlying signal structure remains fundamentally detectable.

\section{Conclusion and Outlook}
We have presented a quantum data augmentation algorithm grounded in quantum harmonic analysis, which can be implemented on a quantum computer with only quadratic depth (assuming access to state preparation oracles). We have demonstrated its efficacy for signal denoising within the small-sample regime. Unlike traditional methods that increase model expressivity through additional layers or kernels, our approach transforms the data itself via a stochastic channel. This data-centric paradigm complements existing research by establishing operator-theoretic methods as a rigorous alternative to purely variational techniques in quantum machine learning.

Future work includes applying the framework to real-world datasets, robustness studies under realistic noise models, and extensions to broader classes of harmonic transforms. Ultimately, we hope that this work stimulates further interaction between quantum harmonic analysis and quantum machine learning, fostering a richer theoretical foundation for data processing on quantum devices.

\section*{Acknowledgment} 
We would like to thank Peter Balazs, Lirand{\"e} Pira, and Erling Arnold T{\o}nseth Svela for inspiring discussions. The authors acknowledge financial support from the Austrian Science Fund via the SFB BeyondC (Grant No.\ F7107-N38), and the START project q-shadows, as well as the European Research Council (ERC Grant Agreement No.\ 101117138, q-shadows).

\section*{Code availability}
The code used to generate the results in this study is publicly available at \url{https://github.com/QUICK-JKU/harmoniq}, including scripts for data generation, analysis, and figure reproduction, as well as the data itself.

\bibliographystyle{IEEEtran}
\bibliography{references}

\end{document}